\documentclass{article}
\usepackage{spconf,amsmath,graphicx}
\usepackage{comment}

\usepackage{enumitem}
\setlist{nosep, leftmargin=14pt}

\usepackage{mwe} 

\usepackage{booktabs}
\usepackage{algorithm,algpseudocode}
\usepackage{blindtext}
\usepackage{hyperref}
\usepackage[table]{xcolor}
\usepackage{graphicx}
\usepackage{amsmath}
\usepackage{booktabs}
\usepackage{overpic}
\usepackage{bbding}
\usepackage{pifont}
\usepackage{wasysym}
\usepackage{amssymb}

\def\ourmodel{DiffUltra}
\definecolor{ForestGreen}{RGB}{34,139,34}
\definecolor{mygray}{gray}{.92}

\algnewcommand{\Inputs}[1]{%
  \State \textbf{Inputs:}
  \Statex \hspace*{\algorithmicindent}\parbox[t]{.8\linewidth}{\raggedright #1}
}
\algnewcommand{\Initialize}[1]{%
  \State \textbf{Initialize:}
  \Statex \hspace*{\algorithmicindent}\parbox[t]{.8\linewidth}{\raggedright #1}
}

\newcommand{\figref}[1]{Figure \ref{#1}}
\newcommand{\tabref}[1]{Table \ref{#1}}

\newcommand{\gary}[1]{{\textcolor{ForestGreen}{\fontfamily{ppl}\selectfont{\textit{#1}}}}}

\title{Ultrasound Image Synthesis using Generative AI \\
for Lung Consolidation Detection}

\name{%
\begin{tabular}{@{}c@{}}
Yu-Cheng Chou$^{\dagger1}$ \qquad 
Gary Y. Li$^{*2}$ \qquad 
Li Chen$^{2}$ \qquad
Mohsen Zahiri$^{2}$ \qquad
Naveen Balaraju$^{2}$ \\ 
Shubham Patil$^{2}$ \qquad
Bryson Hicks$^{3}$ \qquad
Nikolai Schnittke$^{3}$ \qquad 
David O. Kessler$^{4}$ \qquad 
Jeffrey Shupp$^{5}$ \\
Maria Parker$^{3}$ \qquad
Cristiana Baloescu$^{6}$ \qquad
Christopher Moore$^{6}$ \qquad 
Cynthia Gregory$^{3}$ \\ 
Kenton Gregory$^{3}$ \qquad
Balasundar Raju$^{2}$ \qquad
Jochen Kruecker$^{2}$ \qquad
Alvin Chen$^{2}$ \\
\end{tabular}
\thanks{\\
$\dagger$ Work completed during internship at Philips Ultrasound. \\
*Corresponding author: \href{mailto:ye.li@philips.com}{ye.li@philips.com}\\
}}

\address{$^{1}$ Johns Hopkins University   $^{2}$Philips Ultrasound $^{3}$Oregon Health \& Science University \\ $^{4}$Columbia University Vagelos College of Physicians and Surgeons \\ $^{5}$MedStar Washington Hospital Center  $^{6}$Yale University School of Medicine}

\begin{document}
\maketitle
%
\begin{abstract}
Developing reliable healthcare AI models requires training with representative and diverse data.
In imbalanced datasets, model performance tends to plateau on the more prevalent classes while remaining low on less common cases.
%
To overcome this limitation, we propose \ourmodel, the first generative AI technique capable of synthesizing realistic Lung Ultrasound (LUS) images with extensive lesion variability.     

Specifically, we condition the generative AI by the introduced  \textit{Lesion-anatomy Bank}, which captures the lesion's structural and positional properties from real patient data to guide the image synthesis.
We demonstrate that \ourmodel~improves consolidation detection by 5.6\% in AP compared to the models trained solely on real patient data.
More importantly, \ourmodel~increases data diversity and prevalence of rare cases, leading to a 25\% AP improvement in detecting rare instances such as large lung consolidations, which make up only 10\% of the dataset.
\end{abstract}

\begin{keywords}
Synthetic data training, conditional diffusion model, lung consolidation, Video object detection 
\end{keywords}
\section{Introduction}\label{sec:intro}
Recent advancements in generative models have significantly improved the data synthesis for assisting AI training~\cite{li2023early,peng2024optimizing,zhu2024generative,liang2022sketch}.
Typically, the generative models adhere to the \textit{mask-and-paste} generation paradigm, synthesizing only the lesion area guided by pixel-level annotation of target lesion (e.g., segmentation mask) and then pasting the synthesized lesion onto a healthy background~\cite{hu2023label}.
%
However, since the segmentation masks do not capture the internal structure and texture of the target lesion, the synthetic lesions often exhibit uniform structures, making them easily distinguishable from the surrounding tissue by the differences in boundary intensity.~\cite{chen2024towards,wu2024freetumor}.
%

\begin{figure}[!t]
  \centering
  \includegraphics[width=\linewidth]{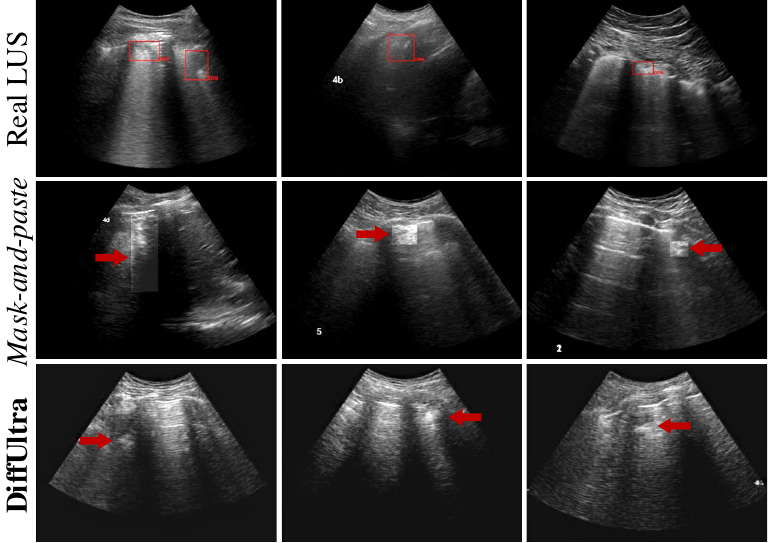}
  \caption{
  Without a pixel-level lesion segmentation mask, the current \textit{mask-and-paste} generation paradigm~\cite{chen2024towards,wu2024freetumor} produces noticeable boundary artifacts, creating a clear distinction between the synthetic lesion and its background (middle row). 
  }\label{fig:teaser}
\end{figure}

Moreover, the \textit{mask-and-paste} generation paradigm becomes impractical when segmentation mask of a lesion is not available, rendering obvious boundary artifacts between the synthetic lesion and its background (\figref{fig:teaser}).
Therefore, in this paper, we aim to move beyond the \textit{mask-and-paste} generation paradigm and investigate a new approach that can synthesize structurally and positionally realistic lesions.

We hypothesize that the uniform structure of synthetic lesions stems from insufficient guidance, such as the conditions \cite{chen2024towards,zhang2024lefusion} used in conditional diffusion models ~\cite{rombach2022high,zhang2023adding}, and poor modeling of the lesion's internal structure ~\cite{li2023early,hu2023label,lai2024pixel}. 
For LUS images in particular, we further hypothesize that lesion location plays a critical role in synthesizing realistic images. Specifically, as shown in ~\figref{fig:module}-(a), lesions of certain sizes and textures can only appear on certain background patterns within the image. Placing lesions at random locations can lead to unnatural blending with the background, making the generated image appear overly artificial and potentially less effective for enhancing the downstream task.


To address this, we propose DiffUltra, a framework for synthesizing whole LUS images with lesions using structural and positional guidance. DiffUltra has two key advantages over existing methods ~\cite{chen2024towards,wu2024freetumor,zhang2024lefusion, li2015development}: (1) it models lesion-to-anatomy positions and internal structures for realistic lesion placement and texture, and (2) it generates full LUS images with only bounding box annotations, reducing annotation effort.
Our experiments on a large dataset demonstrate that (1) DiffUltra produces LUS images with realistic lesion structure and position, (2) synthetic data from DiffUltra improves lung consolidation detection over models trained on real data alone(\S\ref{sec:detecting}), and (3) it outperforms binary conditions (\S\ref{sec:ablation}) by incorporating detailed structural representations. To our knowledge, this is the first approach to synthesize LUS images with this level of realism. Our main contributions are:
\begin{enumerate}
    \item We introduce \ourmodel, a method for synthesizing whole LUS images with realistic lesion structure and position, without requiring pixel-level lesion segmentation.
    %
    \item We show that \ourmodel~improves model reliability for lung consolidation detection compared to the model trained with real data (\tabref{tab:detection}), especially for the rare cases (consolidations of severity level 1 and 4, \tabref{tab:rare}).
    \item We demonstrate that conditioning the generation on structural representations enhances model performance compared to binary conditions (\tabref{tab:abla}).
    \item We show that simply duplicating rare cases does not add new information and fails to improve downstream task performance  (\tabref{tab:parameter}).
\end{enumerate}

\section{\ourmodel}

\begin{figure}[!t]
  \centering
  \includegraphics[width=\linewidth]{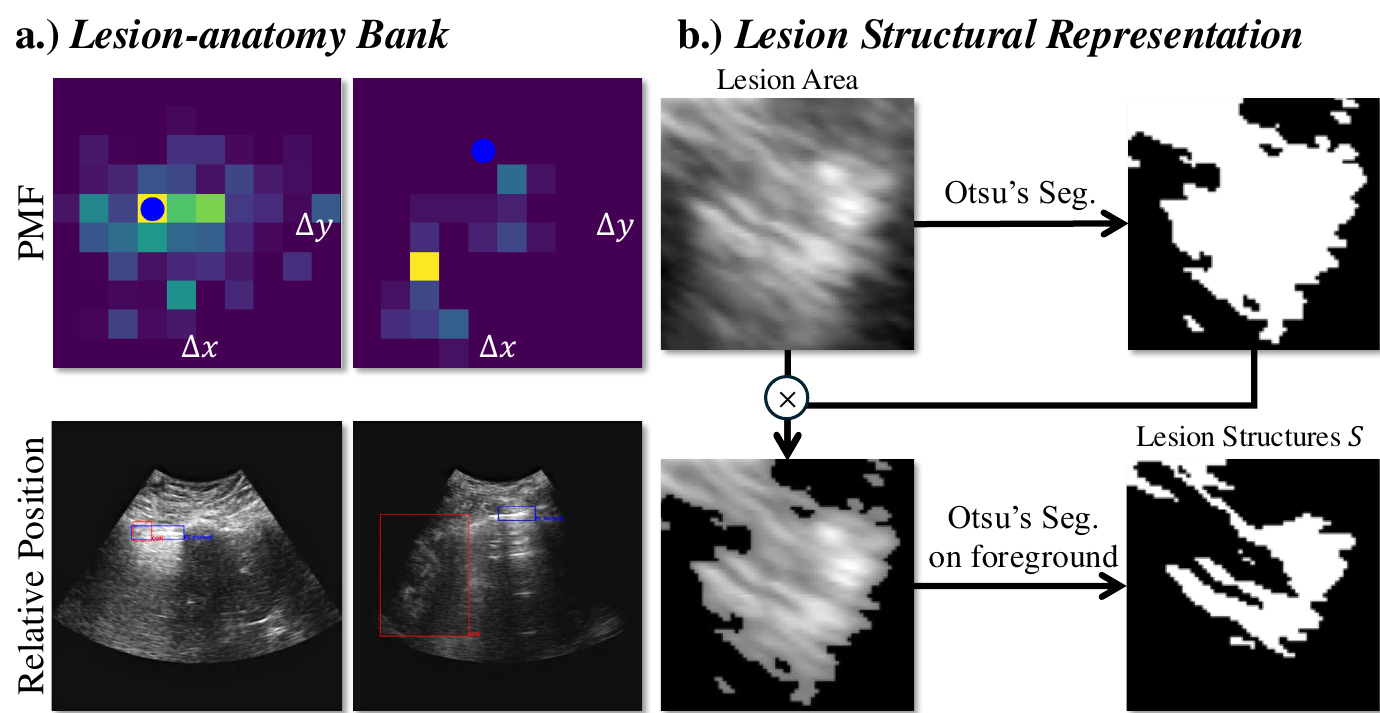}
  \caption{
  a.) The visualization of $P(\Delta X, \Delta Y \mid X=x, Y=y)$, where the blue dot ($x,y$) denotes position of the anatomical structure, and b.) the pipeline for creating the lesion "skeleton"  which is used as condition to guide the generative model.
  }\label{fig:module}
\end{figure}

DiffUltra aims to generate realistic lesions that blend seamlessly into healthy LUS images. To place lesions in anatomically appropriate locations, we use a lesion-anatomy bank, capturing the lesion’s relative position to surrounding structures. 
The lesion-anatomy bank is constructed by creating a joint conditional probability mass function (PMF) and a lesion bank of foregrounds, both derived from real patient data.
For realistic texture, we condition the generative model with a detailed structural representation of the lesion, enabling the synthesis of whole LUS images where lesions integrate naturally with the background anatomy (\figref{fig:module}-(a)).

%

%

\begin{figure*}[!t]
  \centering
  \includegraphics[width=\linewidth]{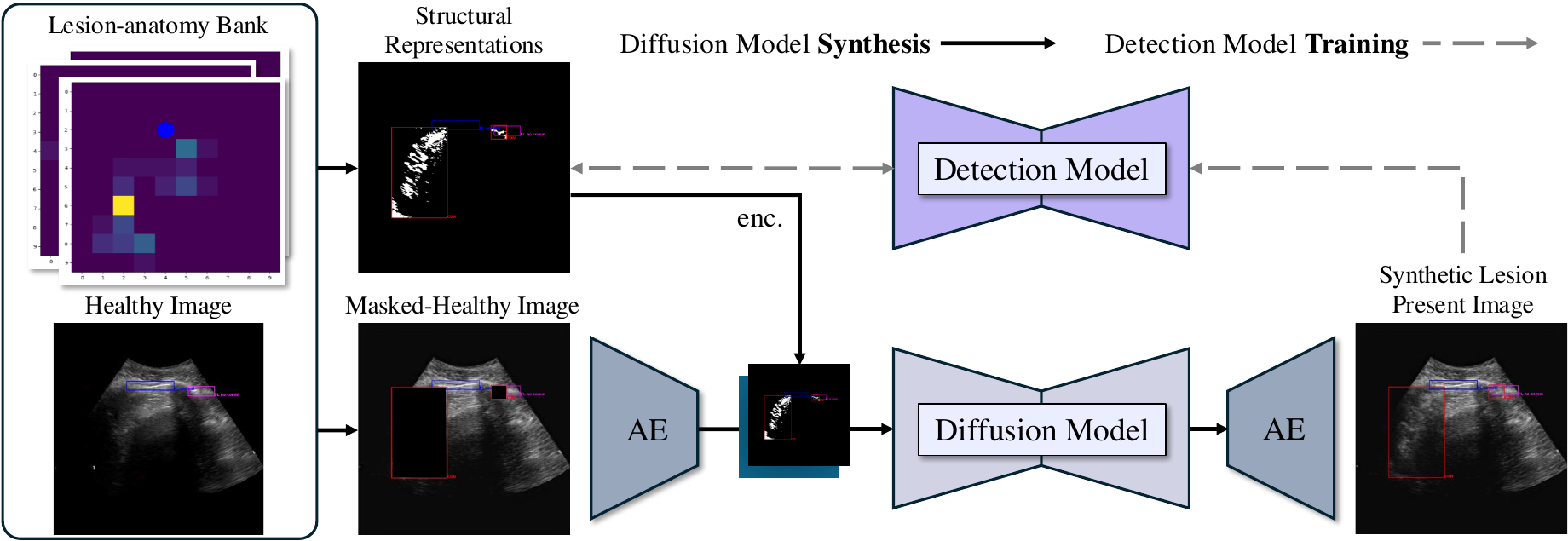}
  \caption{
  The pipeline of the proposed \ourmodel.
  }\label{fig:method}
\end{figure*}

\subsection{Lesion-anatomy Bank}



\subsubsection{Determining appropriate lesion position for synthesis}
To ensure synthesized lesions are placed appropriately relative to their surrounding anatomical structures (e.g., the pleural line) in LUS images, we model the lesion’s relative position to its surrounding anatomical structures  using a conditional PMF - $P(\Delta X, \Delta Y \mid X, Y)$,  built from real patient data. Here, $X$ and $Y$ represent the coordinates of a key anatomical structure’s center, while $\Delta X$ and $\Delta Y$ denote the relative distance between the key anatomical structure and the lesion. This conditional PMF allows us to determine the position of the synthesized lesion by sampling from $P(\Delta X, \Delta Y \mid X=x, Y=y)$, where $x$ and $y$ are derived from a healthy image during synthesis.\figref{fig:module}-(a) visualizes one of these conditional PMF.

To construct the conditional PMF, we first compute the joint PMF $P(\Delta X=\Delta x, \Delta Y=\Delta y, X=x, Y=y)$ using real patient data annotated at the bounding box level. For each lesion, the distance to its nearest key anatomical structure is calculated as:
\begin{equation}
\begin{split}
(\Delta x_i, \Delta y_i) &= (x' - x_i, y' - y_i), \\
(\Delta x, \Delta y) &= \min \left( \sqrt{\Delta x_i^2 + \Delta y_i^2} \right),
\end{split}
\end{equation}

\noindent where $(x', y')$ and $(x_i, y_i)$ are the bounding box centers of the lesion and its surrounding anatomical structure $i$, respectively. This approach allows precise modeling of relative positions, even when multiple key anatomical structures are present in the image.
For each scanning zone and orientation, the joint PMF is built by counting occurrences in a 4D grid (e.g., a $10 \times 10\times10\times10$ grid for a given coordinate system). Next, we obtain $P(X, Y)$ by marginalizing out $\Delta X$ and $\Delta Y$ from the joint PMF. The final conditional PMF $P(\Delta X, \Delta Y \mid X, Y)$ is then obtained by dividing the joint PMF by $P(X, Y)$.

\subsubsection{Selecting appropriate lesion for synthesis}
After determining the position of the lesion to be synthesized in the healthy image (by sampling $P(\Delta X, \Delta Y \mid X=x, Y=y)$), the next step is to select a lesion with the appropriate size and texture that fits the sampled relative position $(\Delta x, \Delta y)$. To achieve this, we propose a lesion bank that stores lesion foregrounds (regions inside the lesion's bounding box) extracted from real patient data, indexed by $(\Delta x, \Delta y, x, y)$.
During inference, a lesion foreground is randomly selected from the bank for the target position. Texture and size information are extracted from the selected foreground using Otsu' segmentation ~\cite{otsu1975threshold} (as shown in~\figref{fig:module}-(b)) and used as conditions for the generative model as shown in~\figref{fig:method}~(\S\ref{sec:diff}).
This process is represented mathematically as:
\begin{equation}
    P(L \mid \Delta X=\Delta x, \Delta Y=\Delta y, X=x, Y=y),
\end{equation}

\noindent where $L$ is the lesion foreground index. Since lesion foregrounds are unique, this conditional PMF is uniform, allowing for random sampling to retrieve a variety of lesion foregrounds for image synthesis.

\subsection{Lesion Structural Representation}
A simple method to add synthetic lesions to healthy images involves pasting a sampled lesion onto a chosen location. However, this method lacks texture variation, as it merely replicates the original lesion.
To enhance variability, we use a generative model conditioned on a detailed structural representation of the lesion—its ``skeleton'' without texture. This approach allows the generative model to introduce texture variation during synthesis.
To extract the lesion skeleton $S$, as illustrated in~\figref{fig:module}-(b), we applied Otsu’s segmentation~\cite{otsu1975threshold} to the lesion foreground capture fine-grained structural details.
\label{lesionSkeleton}

\subsection{Conditional Diffusion Model}\label{sec:diff}
Unlike Medfusion~\cite{muller2023multimodal}, which uses covariables like age and sex, we condition the model on structural representations and latent features. Following ~\cite{muller2023multimodal}, we use a stable diffusion model \cite{rombach2022high} conditioned by structural representations and latent features from a pretrained autoencoder as shown in~\figref{fig:method}. 
With the autoencoder $Dec(Enc(\cdot))$ and diffusion model $D(\cdot)$, we generate lesions in healthy LUS images $\hat{I}$ as:
\begin{equation}
\hat{I} = Dec(D(f, S)),
\end{equation}
where $S$ is the lesion skeleton in \ref{lesionSkeleton} and $f$ is the latent feature, obtained by:
\begin{equation}
f = Enc(\hat{I}_{\text{masked}}),~
\hat{I}_{\text{masked}} = \begin{cases} 
\hat{I}(x, y), & \text{if}~(x, y) \notin F, \\
0, & \text{if}~(x, y) \in F,
\end{cases}
\end{equation}

\noindent with $\hat{I}_{\text{masked}}$ representing the masked healthy image by the entire bounding box area of the lesion foreground $F$, sampled from $P(L \mid \Delta X=\Delta x, \Delta Y=\Delta y, X=x, Y=y)$.
The lesion's center is determined by:
\begin{equation}
(x', y') = (x + \Delta x, y + \Delta y).
\end{equation}

\section{Experiments}

\subsection{Experimental Setting}

\noindent\textbf{\textit{Implementation Details.}}
To reduce the input dimensions for the stable diffusion model, we trained an autoencoder (AE) to downsample LUS frames from $512\times512\times1$  to $64\times64\times8$ latent features. The AE was trained on all frames, and the best checkpoint was selected based on the lowest validation Mean Squred Error (MSE) loss. The diffusion model was then trained in this latent space. The diffusion models were trained for 50 epochs on 4 A100 GPUs with a batch size of 8, and the best checkpoint was chosen based on the lowest MSE loss in the foreground. During synthesis, we randomly sampled healthy images that have pleural line boxes to generate lesion-present images. Following~\cite{chen2024towards}, we replaced the sampled healthy images with synthetic lesion-present images to maintain a consistent number of images in the train set for detection model training. The diffusion model was set to 150 steps during inference. We used YOLO-v5~\cite{Jocher_YOLOv5_by_Ultralytics_2020} and trained it for 300 epochs with default settings.

\noindent\textbf{\textit{Dataset.}}
A dataset of 7,017 LUS videos from 424 patients across 11 U.S. sites, suspected of having lung consolidation, was used. It was divided into training, validation, and testing sets, consisting of 4,930, 1,051, and 599 videos, respectively. Lung consolidation was annotated by Ultrasound-trained physicians, resulting in 45,210 bounding boxes. Additionally, 26,556 images in the training set have annotated pleural lines. Videos were sampled at every 5th frame to improve training efficiency.

\subsection{Experimental Results}
\label{sec:detecting}
\noindent\textbf{Synthetic Data Improves Downstream Tasks}
To evaluate the performance gain provided by \ourmodel, we compare a detection model trained on both real and synthetic data with one trained only on real data.
Results in~\tabref{tab:detection} show that the model trained on both real and synthetic data outperforms the baseline in lesion-level AP and video-level AUROC (+5.6\% and +1.4\%). Furthermore, \ourmodel~outperforms DiffTumor~\cite{chen2024towards} that generates lesions using the \textit{mask-and-paste} paradigm, highlighting the effectiveness of our method in synthesizing complete LUS images.

\noindent\textbf{Synthetic Data Alleviates Class Imbalance}\label{sec:detecting}
To evaluate \ourmodel's impact on improving performance in the low prevalence cases, we conducted a sub-analysis on video-level classification (VLC) of consolidations across four severity levels. To ensure a fair comparison, we matched the VLC specificity of the baseline Yolo-v5 model (89.1\%) with that of \ourmodel~(88.8\%) by adjusting the threshold.
As shown in~\tabref{tab:rare}, \ourmodel~significantly enhances VLC sensitivity for severity level-1 and level-4 consolidations (+22.3\% and +25\%), highlighting its effectiveness in handling rare cases, which represent only 1.5\% and 10.7\% of the testing set, respectively.

\begin{table}[t!]
  \centering
  \scriptsize
  \renewcommand{\arraystretch}{1.15}
  \setlength\tabcolsep{5.2pt}
  \caption{
  Compared to the baseline model trained solely on real data, \ourmodel~can generally improve consolidation detection performance.
  }
  \begin{tabular}{l||rr} 
  \toprule
 & Lesion-level (AP@0.5) & Video-level (AUROC) \\
  \hline
  Yolo-v5 w/o Image Synthesis   &12.7\% &90.0\% \\
  Yolo-v5 + DiffTumor~\cite{chen2024towards}  &14.7\% &91.0\% \\
  \rowcolor{mygray}
  Yolo-v5 + \ourmodel & (+5.6\%) \textbf{18.3}\% &(+1.4\%) \textbf{91.4}\% \\
  \bottomrule
  \end{tabular}
  \label{tab:detection}
\end{table}

\begin{table}[t!]
  \centering
  \scriptsize
  \renewcommand{\arraystretch}{1.15}
  \setlength\tabcolsep{3.4pt}
  \caption{ 
  \ourmodel~ increases data diversity and thereby greatly improves the video-level classification sensitivity for rare cases compared to baseline model, measured at a fixed specificity of 89\%. $Pl$ = Prevalence.
  }
  \begin{tabular}{l||rrrr} 
  \toprule
  & Severity Level 1 &Severity Level 4 &Severity Level 3 & Severity Level 2  \\
  & (\textbf{$Pl$ = 1.5\%}) &(\textbf{$Pl$ = 10.7\%}) &($Pl$ = 15.0\%) &($Pl$ = 23.7\%) \\
  \hline
  Baseline &44.4\% &42.2\% &71.1\% &\textbf{84.5}\% \\
  \rowcolor{mygray}
  \ourmodel &(+22.3\%) \textbf{66.7}\% &(+25\%) \textbf{67.2}\% &(+2.2\%) \textbf{73.3}\% &82.4\% \\
  \bottomrule
  \end{tabular}
  \label{tab:rare}
\end{table}

\subsection{Ablation Study}\label{sec:ablation}

\noindent\textbf{\textit{Excluding structural representation}}.
Conditioning the generative model on a binary mask significantly reduces performance compared to using structural representations (11.7\% vs. 18.3\%, ~\tabref{tab:abla}).
This results are in line with our hypothesis about the need for more detailed structural representation as diffusion condition.

%

\noindent\textbf{\textit{Excluding positional guidance}}.
%
%
We show that randomly placing lesions creates unrealistic relations with surrounding anatomy, reducing detection performance (14.8\% vs. 18.3\%, ~\tabref{tab:abla}). This results are in line with our hypothesis about need for realistic lesion
location during synthesizing.

\noindent\textbf{\textit{Repeating rare cases}}.
We also tested whether simply repeating rare cases would improve performance by balancing the data. However, as shown in ~\tabref{tab:parameter}, repeating rare cases did not enhance performance, indicating that balancing alone, without new information, is insufficient for improvement.
%

\noindent\textbf{Searching Optimal Amount of Synthetic Data}
We optimized the amount of synthetic data generated through experimentation. By replacing only healthy LUS images with lesion-present ones, the overall training data volume stayed constant while increasing the positive-to-negative ($P:N$) ratio. Experimental results are shown in \tabref{tab:parameter}. 


\begin{table}[t!]
  \centering
  \scriptsize
  \renewcommand{\arraystretch}{1.15}
  \setlength\tabcolsep{10.7 pt}
  \caption{ 
  Ablation results in excluding structural representation ($S$) and positional guidance ($PMF$), respectively.
  }
  \begin{tabular}{cc||cc} 
  \toprule
  $S$ & $PMF$ & Lesion-level (AP@0.5) & Video-level (AUROC) \\  
  \hline
   & &13.5\% &89.8\%\\
   &\checkmark &11.7\% &\textbf{91.9}\%\\
  \checkmark &  &14.8\% &91.0\%\\
  \rowcolor{mygray}
  \checkmark & \checkmark &\textbf{18.3}\% &91.4\% \\
  \bottomrule
  \end{tabular}
  \label{tab:abla}
\end{table}


\begin{table}[t!]
  \centering
  \scriptsize
  \renewcommand{\arraystretch}{1.15}
  \setlength\tabcolsep{8.5pt}
  \caption{ 
  Hyper-parameter searching experiments of changing the positive-to-negative ratio ($P: N$) and of simply repeating rare cases ($Repeat$).
  }
  \begin{tabular}{ll||cc} 
  \toprule
  & $P: N$ & Lesion-level (AP@0.5) & Video-level (AUROC) \\  
  \hline
  \hline
  Baseline &1: 7.6  &12.7\% &90.0\% \\
  $Repeat$ &1: 2.3 & 12.7\%  & 92.3\%\\
  \hline
  \ourmodel &1: 3.3 &10.6\%  &91.2\%\\
  \ourmodel &1: 2.5 &13.0\%  &\textbf{92.4}\%\\
  \rowcolor{mygray}
  \ourmodel &1: 1.9 &\textbf{18.3}\% &91.4\%\\
  \ourmodel &1: 1.5 &14.6\%  &91.5\%\\
  
  \bottomrule
  \end{tabular}
  \label{tab:parameter}
\end{table}

\section{Conclusion}

We present DiffUltra, a method for synthesizing Lung Ultrasound images with flexible, clinically accurate lesions. Using a Lesion-Anatomy Bank, DiffUltra captures structural and positional relationships, generating realistic anatomy-lesion combinations to boost data diversity and prevalence. This approach enhances AI detection models, particularly in rare positive cases, and improves AI-based lung consolidation detection and ultrasound diagnostic reliability.

\section{Acknowledgements}
 The study described in this paper was funded in part with federal funds from the U.S. Department of Health and Human Services (HHS); Administration for Strategic Preparedness and Response (ASPR); Biomedical Advanced Research and Development Authority (BARDA), under contract number 75A50120C00097. The contract and federal funding are not an endorsement of the study results, product or company.
\bibliographystyle{IEEEbib}
\bibliography{strings,refs}

\end{document}